\documentclass[12pt,preprint]{aastex}
\begin{document}

\title{Radial Velocity Variability of Field Brown Dwarfs}

\author{L. Prato\altaffilmark{1}, G. N. Mace\altaffilmark{2}, E. L. Rice\altaffilmark{3,4}, I. S. McLean\altaffilmark{5},
J. Davy Kirkpatrick\altaffilmark{6}, A. J. Burgasser\altaffilmark{7}, Sungsoo S. Kim\altaffilmark{8}}

\altaffiltext{1}{Lowell Observatory, 1400 West Mars Hill, Road, Flagstaff, AZ 86001, USA; lprato@lowell.edu}
\altaffiltext{2}{Department of Astronomy, University of Texas, R.L. Moore Hall, Austin, TX 78712, USA}
\altaffiltext{3}{Department of Engineering Science and Physics, College of Staten Island,
2800 Victory Boulevard, Staten Island, NY 10314, USA}
\altaffiltext{4}{Department of Astrophysics, American Museum of Natural History, Central Park
West at 79th Street, New York, NY 10024, USA}
\altaffiltext{5}{Department of Physics and Astronomy, UCLA, 430 Portola Plaza, Box 951547, Los Angeles, CA 90095-1547, USA}
\altaffiltext{6}{Infrared Processing and Analysis Center, California Institute of
Technology, Pasadena, CA 91125, USA}
\altaffiltext{7}{Center for Astrophysics and Space Science, University of California San Diego, La Jolla, CA 92093, USA}
\altaffiltext{8}{Department of Astronomy \& Space Science, Kyung Hee University, Yongin, Kyungki 446-701, Republic of Korea}

\begin{abstract}

We present paper six of the NIRSPEC Brown Dwarf Spectroscopic Survey, an analysis of multi-epoch, high-resolution (R$\sim$20,000)
spectra of 25 field dwarf systems (3 late-type M dwarfs, 16 L dwarfs, and 6 T dwarfs) taken with the NIRSPEC infrared spectrograph
at the W. M. Keck Observatory.  With a radial velocity precision of $\sim$2 km~s$^{-1}$, we are sensitive to brown dwarf companions
in orbits with periods of a few years or less given a mass ratio of 0.5 or greater.  We do not detect any spectroscopic binary brown
dwarfs in the sample.  Given our target properties, and the frequency and cadence of observations, we use a
Monte Carlo simulation to determine the detection probability of our sample.  Even with a null detection result, our 1 $\sigma$ upper
limit for very low mass binary frequency is 18\%.
Our targets included 7 known, wide brown dwarf binary systems.  No significant radial velocity variability was
measured in our multi-epoch observations of these systems, even for those pairs for which our data spanned a significant fraction
of the orbital period.  Specialized techniques are required to reach the high precisions sensitive to motion in orbits of very low-mass systems.
For eight objects, including six T dwarfs, we present the first published high-resolution spectra,
many with high signal to noise, that will provide valuable comparison data for models of brown dwarf atmospheres.

\end{abstract}


% keep this FYI:  "The mass-ratio distribution shows a preference for like-mass
% pairs, which occur more frequently in relatively close pairs."
% (from abstract of Raghaven et al. 2010 -- but keep in mind that this refers
% to F6 - K3 stars.)

\section{Introduction}

Star and planet formation processes both give rise to objects in the $\sim$1 to 20 M$_{Jup}$ range
(e.g., Luhman et al. 2005; Bakos et al. 2009; Marois et al. 2010; Joergens et al. 2013).
Naively, objects of higher mass are typically assumed to form primarily via the star formation process and
objects of lower mass are assumed to form primarily from a proto-planetary disk.  This
simplification is directly testable with a variety of approaches, both
theoretical and observational.  One straightforward observational approach
is the study of multiplicity: do brown dwarfs come in pairs as frequently and
with the same binary properties as stars?  Although trends are suggestive
among stellar visual binaries, i.e. decreasing frequency with later spectral types
(e.g., Siegler et al. 2003), the frequency of short-period brown dwarf binaries is
relatively uncertain (Burgasser et al. 2012).

Binary parameter space is multi-dimensional.  For a given 
spectral type, binaries may be examined in terms of companion frequency, separation,
mass ratio distribution, or secondary mass distribution.  Mazeh et al. (2003),
for example, showed that for main sequence M stars the secondary mass distribution
does not conform to a standard initial mass function (IMF) but instead follows a
relatively flat distribution for a primary sample
with M$\sim$0.7 $\pm$0.1 M$_{\odot}$ and orbital period P$<$3,000 days.
This represents one small slice of a parameter space that may also be studied
in diverse populations:  young, intermediate age, old,
metal-poor, clustered, etc.  Specific comparisons between these samples provide a wealth
of diagnostics for understanding the similarities and differences in formation
and evolution of distinct groups of objects.

In a sample of 454 F6$-$K3 solar-type stellar systems within 25pc of the Sun,
Raghavan et al. (2010) found a total fraction of binaries and higher order
multiples of 44\%\footnote{Thus, although the majority of these {\it systems} may not
be multiple, the majority of the {\it stars} studied reside in multiple systems,
as previously concluded (e.g., Abt \& Levy 1976; Duquennoy \& Mayor 1991).}.
Among other results, they reconfirm the existence of the brown dwarf desert,
the pronounced dearth of brown dwarf mass (i.e. $\sim$10$-$80 M$_{Jup}$) companions
to stars in orbits with periods less than a few years (e.g., Grether \& Lineweaver 2006;
Metchev \& Hillenbrand 2009). 

Is the brown dwarf desert the result of dynamical evolution preferentially
impacting lower mass companions (e.g., Reipurth \& Clarke 2001; Armitage \& Bonnell 2002)
or does it have more to do with poorly understood
barriers to the formation of tightly bound companions of brown dwarf mass?  
In a radial velocity (RV) survey with a few hundred m/s precision of $>$100 stars in the Taurus star
forming region (Crockett et al. 2012), no brown dwarf companions to 1$-$3 Myr old
stars have been observed (Mahmud et al. 2015, in prep), indicating that the existence of the
desert is more likely related to formation than dynamical evolution in
origin.  Among 1$-$10 Myr old populations, to date only one brown dwarf-brown dwarf
short-period ($\la$1 year) spectroscopic binary pair has been identified (Stassun et al. 2006).
Joergens et al. (2010) found an orbital period of 5.2 years
for the young Chameleon brown dwarf binary Cha~H$\alpha$8 and Joergens (2008) estimates a period
of $>$12 years for the pair CHXR74.  However, the results of Stassun et al. and Joergens are
based, respectively, on a survey for eclipsing systems and on a relatively small sample and thus are
likely incomplete.  Joergens estimates a binary frequency among very low mass young objects
of $10^{+18}_{-8}$\%.

Among intermediate age brown dwarf spectroscopic binaries, Basri \& Mart\'{i}n (1999)
found the first brown dwarf pair, PPL 15, a 5.8 day period system, in a study of the
Pleiades.  Simon et al. (2006) studied the 100 Myr old brown dwarf
binary GL~569B (Zapatero Osorio et al. 2004),
a pair with a $\sim$2.4 year orbit, a semi-major axis of 0.89~AU.
They detected some evidence for a third spectroscopic component in the system, yet to be confirmed.
%{\bf LP349-25 (Forveille et al. 2005), an M8$+$M9 system with an age of 140 Myr (Dupuy et al. 2010),
%shows RV variability consistent with a $\sim$7.3 year orbit with a semi-major axis of $\sim$1 AU
%(Konopacky et al. 2010).}

RV surveys among field brown dwarfs, sensitive to binaries with periods
of several years or less and semi-major axes of a few AU, have yielded a handful of definitive detections. 
In a sample of 59 field brown dwarfs, Blake et al. (2010) found a tight binary (a$<$1~AU) frequency
of $2.5^{+8.6}_{-1.6}$\%. They had previously identified and measured the orbit of the $\sim$247 day
period system 2MASS 0320-04 (Blake et al. 2008), independently identified on the basis of spectral analysis by Burgasser et al. (2008),
combined their RV measurements with the astrometry of Dahn et al. (2008)
for the $\sim$607 day period system LSR J1610-0040, and presented RV data on two wide
substellar pairs with periods of $>$10 years (2MASS J15074769-1627386 and 2MASS J07464256+2000321).
Zapatero Osorio et al. (2007) measured space motions for over a dozen field
brown dwarfs but found no spectroscopic pairs in their sample.
Burgasser et al. (2012) presented a solution for the spectroscopic orbit of
the 148 day period pair SDSS J000649.16-085246.3AB, in common proper motion with
the very low mass star LP 704-48, with M9 and T0 components straddling the substellar limit.
Although Basri \& Reiners (2006) indicate an overall spectroscopic binary fraction for
field brown dwarfs and very low mass stars of 11\% in their RV survey of 53 targets,
only three L dwarfs in their sample show some level of RV variability.  Of these three,
2MASS J15065441$+$1321060 was subsequently shown by Blake et al. to be non-variable and
2MASS J15074769-1627386 and 2MASS J07464256+2000321 are long-period systems identified as
binaries with imaging observations.  Other brown dwarf pairs have been identified
with imaging (e.g., Lodieu et al. 2007), microlensing (e.g., Bennett et al. 2008), and astrometry
(e.g., Sahlmann et al. 2013).  Spectral brown dwarf binaries, systems that appear single in light of
existing data but are spectroscopically peculiar, implicating the possible presence of a companion, may be
numerous.  Bardalez Gagliuffi et al. (2014) have identified 50 candidates, although Artigau et al. (2009)
and Radigan et al. (2013) identified cases in which the brown dwarf binary candidates were instead found to have
heterogeneous cloud covers.  Thus it is unlikely that all spectral binaries have $<$1 year orbits, but two have been
confirmed as such to date (Blake et al. 2008; Burgasser et al. 2012).

No short period ($<$100 days) field brown dwarf spectroscopic binaries are known, but one
intermediate age and one young system with periods of just a few days were identified by
Basri \& Mart\'{i}n (1999) and Stassun et al. (2006), respectively. Short period systems ought to be
the most straightforward to identify; however,
RV surveys for brown dwarf multiples require the world's biggest
telescopes and generous time allocations, challenging to obtain, and are fraught with bias.  
Yet without such work our understanding of substellar multiplicity is
skewed towards the anecdotal, and astronomers' grasp of the basis for planetary
mass companion formation is isolated from the context of brown dwarf and stellar
mass companion formation.

We report here on 11 years of dynamical observations of over two dozen
field brown dwarfs taken at high spectral resolution in the near-infrared (IR) at the Keck II telescope.  
The intrinsic faintness of brown dwarfs, particularly the late L and T types, presents a challenge to
high-resolution spectroscopic observations.  However, this is the only method by which we may
derive the RV data necessary for calculating space motions, and hence
possible moving group or cluster membership, and the telltale RV variability of a short-period
binary.  Measurements of $v \sin i$ provide a lower limit on rotational velocity, crucial for understanding angular
momentum evolution.  Ultimately, with sufficiently precise data, the combination of RV versus phase together with
the angularly resolved orbits for the few year or few tens of year period systems will yield the
absolute component masses of brown dwarfs in binaries, invaluable for furthering our understanding
of brown dwarf structure and evolution (Konopacky et al. 2010; Dupuy et al. 2010).  Because brown dwarfs emit the
bulk of their energy at wavelengths greater than $\sim$1~$\mu$m, IR spectroscopy provides the
best approach for their RV measurements.

Our goals were to identify brown dwarfs in spectroscopic
binary systems and to measure the dynamical properties of any such pairs
discovered.  This project to identify short-period
brown dwarf multiples is the latest contribution to the NIRSPEC Brown Dwarf Spectroscopic Survey
(BDSS; McLean et al. 2001, 2003, 2007; McGovern et al. 2004; Rice et al. 2010) and leverages over a
decade of observations to characterize brown dwarfs at high spectral resolution.
We find that a critical factor in a productive survey hinges on the RV precision; sensitivity to
RV variability scales rapidly with this parameter. 
We describe our sample, observations, and data reduction in \S 2 and discuss our
data analysis in \S 3.  Section 4 provides a discussion of our results and we briefly
summarize our work in \S 5.

\section{Observations and Data Reduction}

Targets were selected for a range of spectral types across the span of late M, L, and T dwarfs and on the basis of 
magnitude ($J\la15$ mag) and accessibility from the Keck Observatory ($\delta\ga-30^{\circ}$).  
The complete target list and observing log appears in Table 1 which lists the object name (column 1),
Right Ascension and Declination (columns 2 and 3), spectral type (column 4),
$2MASS$ J magnitude (column 5), reference for discovery paper (column 6), and the UT dates of observation (column 7).

Observations were carried out with the high-resolution, cross-dispersed echelle mode
of the facility, near-infrared, cryogenic spectrograph NIRSPEC (McLean et al. 1998, 2000)
on the Keck II 10 m telescope on Mauna Kea, Hawaii. The NIRSPEC 
science detector is a 1024 $\times$ 1024 pixel ALADDIN InSb array; a 256 $\times$ 256
pixel HgCdTe array in a slit viewing camera was used for source acquisition.
The N3 (J-band) filter with the 12 $\times$ 0$\farcs$432 (3-pixel) slit, an echelle angle of 63.00$^{\circ}$, 
and a grating angle of 34.08$^{\circ}$ produces a resolving power
of R = $\lambda$/$\Delta \lambda$ $\approx$ 20,000 and nearly 
continuous coverage from 1.165$-$1.324$\mu$m (orders 58$-$65; McLean et al. 2007).
Observations made on 2000 July 25 and 29 employed the 12 $\times$ 0$\farcs$576 (4-pixel) slit, yielding
a resolution of $\sim$15,000.  Internal white-light spectra, dark frames, and arc lamp spectra were obtained for
flat-fielding, dark current correction, and wavelength calibration.
Science exposures were made in 600~s nodded AB pairs at two locations along the slit.

All spectroscopic reductions were made using the REDSPEC package, software produced
at UCLA by S. Kim, L. Prato, and I. McLean specifically for the analysis of NIRSPEC
data\footnote{See: http://www2.keck.hawaii.edu/inst/nirspec/redspec.html} as 
described in McLean et al. (2007). Wavelength solutions were determined using the OH night
sky emission lines in each order; 4$-$5 OH lines across each of the orders used yielded
wavelength solutions with typical uncertainties of better than 0.4 km~s$^{-1}$.
The two spectral orders most favorable for the analysis, 62 for the L dwarfs (1.221 $\mu$m$-$1.239 $\mu$m; Figure 1)
and 59 for the T dwarfs (1.283 $\mu$m$-$1.302 $\mu$m; Figure 2), were
selected independently on the basis of the presence of deep inherent lines in the brown dwarf targets.
Furthermore, an additional advantage of these particular orders is the absence of terrestrial absorption lines,
thus avoiding the necessity of division by a featureless telluric standard star.
This provided the optimal approach for several reasons:  (1) eliminating division by
telluric standards maximized the signal-to-noise ratio and avoided the possible introduction of
slightly offset spectra and potential small shifts in the brown dwarf absorption lines and hence RV measurements,
(2) focusing on the narrowest and deepest lines available yielded the highest possible RV precision;
although the KI lines in orders 61 and 65 are deep (e.g., McLean et al. 2007), their breadth is unfavorable
to precision RV measurements through cross-correlation, and (3) selecting orders 62 and 59 further
guaranteed the best possible RV precision given the regular spacing of the OH night sky emission lines across both
of these orders, required for a superior dispersion solution; this condition was not met for all orders in our
J band setting.  Multiple-epoch sequences for the L2 dwarf
Kelu-1 and the peculiar T6p dwarf 2M0937 are shown in Figures 3 and 4, respectively.

\section{Analysis}

\subsection{Radial Velocities}

Rice et al. (2010) found typical systematic RV uncertainties of 1$-$2 km~s$^{-1}$ for a sample
observed with a similar methodology, similar signal to noise, and with some overlap in target
data with this paper (Table 2).  We thus adopt
a conservative 2 km~s$^{-1}$ internal uncertainty for our RVs.
We tested this estimate by cross-correlation of the RV invariant target 2M0036 (Blake et al. 2010), an L4 dwarf,
for which we obtained 7 epochs over more than 5 years.
The maximum RV shift between epochs was 1.91 km~s$^{-1}$; the standard deviation
in the RV shift for all epochs was 0.59 km~s$^{-1}$.  Thus 2 km~s$^{-1}$ provides a reasonable
if conservative internal uncertainty on individual RV measurements.  

At least two, and as many as seven, spectra were taken for each of our targets.
We tested for radial velocity variability by cross-correlating the
highest signal-to-noise spectrum against the spectra from all other epochs for a given target;
no significant variability was detected (\S 4).
Table 2 lists the number of spectra taken for each object (column 2) and the
total number of days spanned by the observations (column 3).  RVs
(column 4) were either taken from Blake et al. (2010) or determined by cross-correlation
of the highest signal-to-noise spectrum for a particular object with spectra of objects with known RV
(from Blake et al.) and similar spectral type, sometimes of type both earlier and later than our target.
The RVs resulting from cross-correlation of a target with more than one other object
were averaged and the standard deviation added in
quadrature with the internal uncertainties in the radial velocity measurements.  This result
in most cases was dominated by the 2 km~s$^{-1}$ internal uncertainty; however, for a few objects,
primarily the fainter and thus lower signal to noise late T dwarfs,
this procedure resulted in an uncertainty of 3 km~s$^{-1}$ (Table 2).

We use the average RV values from Blake et al. (2010) when available because of their unprecedented precision,
obtained by fitting models to near-IR K-band CO
bandhead at $\sim$2.3 $\mu$m target spectra.  The models are composed of synthetic
template spectra plus observed telluric spectra; the CO bandhead region of the telluric spectrum is rich in deep
CH$_4$ lines that provide a wavelength dispersion and zero-point reference with a precision as
good as a few tens of m/s.  Small iterations of the RV shift between the synthetic photospheric spectra
and the telluric spectra allow for high accuracy in the target RV measurements.
We compared our results with values from other RV studies in the literature, e.g., Basri et al. (2000),
Mohanty \& Basri (2003), and Zapatero-Osorio et al. (2007).  In every case our RVs were comparable
to other values within 1~$\sigma$.  We provide our results where indicated in Table 2.

\subsection{Rotational Velocities}

Column 5 of Table 2 lists the $v \sin i$ values for our targets.  Most of these were taken
from the literature (Basri et al. 2000; Mohanty \& Basri 2003; Rice et al. 2010; Blake et al. 2010; references given in column 6).  
To estimate $v \sin i$ values for the remaining targets, we used visual comparison with objects
of neighboring spectral types after superimposing the spectra.  For some objects we
convolved comparison spectra of known $v \sin i$ with a boxcar kernel in order to produce 
resulting spectra of larger $v \sin i$ for comparison.  This method was approximate and yielded
uncertainties of 5$-$10 km~s$^{-1}$, based on visual comparisons with objects of known
$v \sin i$, for the T dwarfs in our sample.  Nevertheless, these 
are the first estimates available for some of the targets and thus provide a useful guide.

\section{Discussion}

\subsection{Field Brown Dwarf Spectroscopic Multiplicity}

Konopacky et al. (2010) obtained angularly resolved imaging and spectroscopy for each
component in 24 very low mass stellar and brown dwarf subarcsecond {\it visual} binaries 
contributing to eventual measurements of orbital solutions and
component masses.  Our goal was to use high spectral resolution observations to identify
any RV variability over time that might indicate a {\it spectroscopic} binary brown dwarf.
This requires binaries with orbital periods sufficiently short to measure the component
motion at a significant level, i.e. at least several $\sigma$ above the RV uncertainty.  

To explore our sensitivity to the brown dwarf binary parameter space, given our $\sim$2~ km~s$^{-1}$ RV precision, 
we ran a Monte Carlo simulation of 10$^5$ possible binary orbits for each of the 25 objects in our sample, following
Burgasser et al. (2014).  Orbital parameters were
uniformly distributed in log semi-major axis (10$^{-3}$ to 10$^2$ AU),
mass ratio (0.8$-$1.0), sine inclination (0$-$1), eccentricity (0$-$0.6; Dupuy \& Liu 2011), and all
other orbital parameters (argument of periapsis, longitude of ascending node, and mean anomaly, 0$-$2$\pi$).
We converted our target spectral types to effective temperature using the empirical relation of Stephens et al. (2009),
and then from effective temperature to mass using the evolutionary models of Burrows et al. (2001), assuming ages of 0.5 Gyr, 1.0 Gyr, and 5 Gyr.  
Each of the 10$^5$ simulated orbits was sampled at the dates given and the primary orbital RV was
calculated.  A binary detection, for a given semi-major axis bin (0.2 dex), was defined as a system for which a maximum RV
difference between all dates was $>$3$\sigma$, i.e. $>$6~km~s$^{-1}$ given our 2~km~s$^{-1}$ precision.
The results are summarized in Figure 5.  The most important factor impacting the probability
of detecting a potential binary was the frequency of observation for a given target (Table 2).

A binary with a separation of $\la$0.1 AU should in principle be straightforward to detect with an RV precision of  2~km~s$^{-1}$.
However, given our estimated target masses and sampling frequency, and assuming an age of 1.0 Gyr,
we could have detected such an orbit only 50\% of the time for only 12 of the sources in
our sample (middle left-hand panel of Figure 5).  The detection probability for an 0.1 AU binary fails to
reach 90\% for {\it any} of our sources.  Using the probabilities of detection for separations greater than
a given threshold $a$, $P(>a)$, as a measure of the effective sample size, $N_{eff}(a) = \Sigma_i P(>a)$,
we find our null result translates into a 1$\sigma$ upper limit of 18\% for spectroscopic
binaries down to $a=0.1$~AU, based on binomial statistics.
Only for systems with separations below 0.01 AU ($\sim$1 day orbits) could the spectroscopic binary frequency
of our sample be characterized as relatively rare, i.e. $\la$10\%.

These limits apply when we consider the detectability of individual systems. However, a signature of unresolved multiplicity could also emerge in higher velocity dispersions for the sample as a whole.  Identifying higher dispersions across the sample requires robust determination of the individual measurement uncertainties, but we can perform a rough assessment as follows. Using the same simulation parameters, we calculated the distribution of velocity dispersions one would obtain if a given fraction of sources (randomly selected) were binaries with semi-major axes in logarithmically spaced bins.  For a sample devoid of binaries, the mean dispersion is somewhat less than the adopted measurement uncertainty, about 1.75~km~s$^{-1}$.  Sources with radial orbital motion drive the mean velocity dispersions of the sample higher. Figure~6 displays the thresholds at which the mean simulated sample velocity dispersions are 1.5, 3 and 5 times higher than the dispersions assuming a 2~km~s$^{-1}$ measurement uncertainty. The most conservative threshold is reached at a semi-major axis of 0.03--0.04~AU, and is detectable at even small binary fractions (i.e., 1--2 sources in the sample being binary). This analysis is roughly consistent with the individual detection limits above, and again implies that we can rule out a significant fraction of binaries ($\gtrsim$10\%) only for separations $\lesssim$0.01~AU.

\subsection{Notes on Known Visual Binaries}

Of the 25 targets in our sample, 7 are known visual binaries.    For these systems we estimated the upper limit for the
observable RV shift, $\Delta$(RV)$_{max}$, for the brighter binary component between two epochs, assuming the
most favorable possible observing conditions:
(1) the epochs correspond to the two phases at which the primary component is moving toward and away from us with
maximum RV, (2) the projected separation corresponds to the semi-major axis
of the system, and (3) the orbit is circular and edge-on.  The observed and estimated binary properties are given
in Table 3.  A discussion of each visual binary and the results of our observations follows.

\subsubsection{2MASS J22344161+4041387  $-$  M6}

Using laser guide star adaptive optics imaging at the Keck II telescope,
Allers et al. (2009) identified 2M2234 as a 1 Myr year old, visual binary with a projected physical separation of
51 AU.  Given the observed binary properties, the $\Delta$(RV)$_{max}$
is $\sim$1.9 km~s$^{-1}$ (Table 3).
With an orbital period of $824^{+510}_{-310}$ years the
inclination is effectively indeterminable.  This estimate for the period is based on a circular orbit.
A more realistic value, $1000^{+1600}_{-500}$ years, is calculated 
in Allers et al. (2009).  In either case, it is not possible to observe the system at phases separated by half the orbit.
Furthermore, given the $v \sin i$ of 17 km~s$^{-1}$ for 2M2234 (Table 2), it is also impossible
to spectroscopically resolve the RVs of the two components in a single epoch spectrum, even though the
component near-IR magnitudes are almost equal, because the maximum relative component RV separation is
significantly less than the rotational broadening.

Cross-correlation of our 3 epochs of spectra with each other demonstrated no RV shift between
the 2007 and 2009 data.  Between the 2006 and 2007 data there was an apparent shift of
$-$7.7 km~s$^{-1}$; however, the signal to noise of the 2006 spectrum ($\sim$20) is considerably
lower than that of the other epochs ($\sim$80), and the spectra are veiled (Allers et al. 2009),
thus we do not have confidence in the 2006 result.
Using the young M6 brown dwarf [GY92] 5 for cross-correlation with our 2007 spectrum, we
obtain an RV of $-$10$\pm$2 km~s$^{-1}$ (Table 2)\footnote{Cross-correlating the same 2M2234 spectrum with another young M6,
CFHT Tau 7, Rice et al. (2010) found $-$13.1~km~s$^{-1}$.}, similar to the results of Allers et al. on the basis 
of Keck HIRES data from 2006\footnote{This result is the weighted mean of two RVs;
Shkolnik et al. (2012) use the same Keck
measurements to determine an unweighted mean of $-$10.9$\pm$0.7 km~s$^{-1}$.}, $-$10.6$\pm$0.5 km~s$^{-1}$.
Allers et al. raise the possibility that 2M2234 could be a higher
order multiple system, which would account for the overluminous nature of the A component.
Our multi-epoch observations failed to detect any short-period, i.e. P $<$a few years, hierarchical spectroscopic binary in this system,
although our sensitivity to intermediate separation binaries, and binary orientations unfavorable for detection, limit
any significant statistical conclusions (\S 4.1).  Given the
greater $K_s-L'$ excess in the 2M2234A, it is feasible that the excess luminosity is related to the
circumstellar disk structure, orientation, and/or possible accretion activity.  Such a mismatch in disk properties around
the components of very low mass binaries is not unprecedented; for example,
the TWA 30AB wide, co-moving pair has an apparently edge-on disk around the embedded, earlier-type component, extinguishing this 
late type M star by 5 magnitudes with respect to the cooler component (Looper et al. 2010).

\subsubsection{2MASS J07464256+2000321  $-$  L1}

The 2M0746 binary is a nearby ($d\sim12$ pc), tight ($\sim$3 AU) system.
We use the Konopacky et al. (2010) astrometric measurements (Table 3) to determine a $\Delta$(RV)$_{max}$
of 2.0 km~s$^{-1}$.  Konopacky et al. find an average primary/secondary flux ratio of 1.5 $\pm$0.1, challenging the assumption
that angularly unresolved spectra are fully dominated by the primary (Blake et al. 2010).  

We observed 2M0746 at two epochs separated by almost exactly 4 years, about 1/3 of the orbital period.
Cross-correlation of our two order 62 J-band spectra yielded a 1.3 km~s$^{-1}$ shift with a high correlation coefficient, 0.92.
Comparing the epochs of our observations with the RV curve plotted for this system in Figure 14 of Blake et al. (2010), this is almost
exactly the expected result; however, we are not sufficiently confident in our RV uncertainties to give it much weight.

\subsubsection{Kelu-1   $-$  L2}

A rapid rotator with $v \sin i$ of $\sim$70 km~s$^{-1}$ and Li absorption, Kelu-1 was identified as a brown dwarf by Ruiz et al. (1997).
Mart\'{i}n et al. (1999) hypothesized that Kelu-1's over-luminosity and Li abundance might be explained by a young age or an
additional component in the system (e.g., Golimowski et al. 2004).  
Liu \& Leggett (2005) using Keck AO imaging found that Kelu-1 was a 0$\farcs$291 binary.
Gelino et al. (2006) estimated spectral types for the components of L2 and L3.5 and a total mass of 0.115 $\pm$0.014 M$_{\odot}$.
In an unpublished preprint, Stumpf et al. (2008) describe additional observations of the system with VLT AO imaging
through 2008; the separation steadily increased to 0$\farcs$366 in 2008.  The position angle has not changed by more than
$4^{\circ}$ or $5^{\circ}$.  Adopting the largest separation observed by Gelino et al. as the semi-major axis, 0$\farcs$298 $\pm$0$\farcs$003,
we estimate a period of 39 $\pm$5 years based on a circular orbit (although Stumpf et al. favor a high eccentricity of 0.82 $\pm$0.10).
If viewed edge-on, this implies a $\Delta$(RV)$_{max}$ of 4.3 $\pm$0.4 km~s$^{-1}$, marginally detectable with our $\sim$2 km~s$^{-1}$ precision.

Measurements of the Kelu-1 system RV in the literature are inconsistent:  Basri et al. (2000)
found 17 $\pm$1 km~s$^{-1}$ in June of 1997 and Blake et al. (2010) determined RVs of 6.35 $\pm$0.39 and 6.41 $\pm$0.75 km~s$^{-1}$
in March and April of 2003.  On the basis of angularly resolved spectra of the two
known components, Stumpf et al. (2008) suggest that Kelu-1 A
is itself a spectroscopic binary.  We used our highest signal to noise (S/N) ratio spectrum of Kelu-1 to cross-correlate against five other epochs
(Figure 3), all of S/N ratio $>$50 per resolution element (the January, 2006, spectrum, with a S/N of $\sim$10, was not included in this analysis).  
Our RV measurements, from 2002 through 2011, show RV shifts of $<$3 km~s$^{-1}$.
We did not detect any clear evidence in our spectra for additional motion resulting from the A-component moving
in a relatively short-period spectroscopic orbit; however, this could conceivably be the result of binary properties and/or viewing geometry (\S 4.1).

\subsubsection{2MASS J15074769-1627386  $-$  L5.5}

Over a 6-year baseline,
Blake et al. (2010) detect a marginally significant ($<$2 $\sigma$) trend in the RV of 2M1507, a
nearby (d$=$7.3 pc) L dwarf.  They obtain a false alarm probability of 2.2 \% and suggest the possibility that 2M1507 is a $>$5000 day
binary with an angular separation of 0$\farcs$4.  However, deep, high-resolution imaging sensitive to a
contrast ratio of 5 magnitudes (Bouy et al. 2003; Reid et al. 2006)
has not revealed any companions.  No significant RV variations are evident in the 5 high-resolution spectra we obtained between
2000 and 2008; cross-correlation of the highest S/N ratio spectrum (UT 2000 April 25) against the other 4 epochs resulted in RV
shifts of $<$1.7 km~s$^{-1}$ with an uncertainty of $\sim$2 km~s$^{-1}$.  This result, however, does not rule out multiplicity;
Blake et al. observed $\sim$0.5 km~s$^{-1}$ of motion over 6.5 years, thus we would not expect much more than that over our
8 year baseline.  Given the lack of definitive evidence for multiplicity, this system is not included in Table 3.

\subsubsection{DENIS-P J0205.4-1159  $-$  L5.5}

Koerner et al. (1999) initially identified this system as binary. Bouy et al. (2005) describe evidence for a third
object in a bound orbit with the secondary component.  The estimated properties of the wide binary orbit are uncertain but the period is at least 47 years
and the $\Delta$(RV)$_{max}$ is at most 4.4 km~s$^{-1}$ (Table 3).  For the presumed close binary, Bouy et al. estimate an orbital period of
8 years and a semi-major axis of 1.9 AU, implying a $\Delta$(RV)$_{max}$ of $\sim$7 km~s$^{-1}$.  Our three spectra of DENIS 0205,
taken in 2001 and in 2006, are of low S/N ratio.  Cross-correlation between the epochs yields $-$2.7 and $-$2.1 km~s$^{-1}$
with a correlation coefficient of only $\sim$0.4, reflecting the poor quality of the data.  Sufficiently frequent and deep
imaging and RV monitoring of this system may provide
the requisite phase coverage, preferably with better precision than 2 km~s$^{-1}$, to determine a full orbital solution for
the inner binary over the course of one orbital period.

\subsubsection{DENIS-P J1228.2-1547  $-$  L6} 

Using the {\it Hubble Space Telescope}, Mart\'{i}n et al. (1999) identified DENIS 1228 as the first angularly resolved brown dwarf - brown dwarf
pair with a separation of 0$\farcs$275$\pm$0$\farcs$002 (Bouy et al. 2003).  After several years of monitoring the components' positions,
Brandner et al. (2004) estimated the orbital properties of the system, listed in Table 3.  The $\Delta$(RV)$_{max}$ for this binary, 4.3 km~s$^{-1}$,
in combination with the period of $\sim$44 years from Brandner et al., is not favorable for the detection of an RV shift over the 4 year time scale
of our NIRSPEC observations.  Cross-correlating our 2007 May spectrum with those taken in 2011 February and June yields a
0 km~s$^{-1}$ RV shift.  Continued monitoring of the visual orbit with high angular resolution imaging and high precision RV spectroscopic
techniques will help to refine the parameters
over the next decades, necessary to determine individual component masses in the long term. 

\subsubsection{SDSSp J042348.57-041403.5  $-$  T0}

SDSS 0423 is one of the visual brown dwarf binary systems which spans the L and T classes.  Burgasser et al. (2005) measured a
separation of 0$\farcs$16 and estimated a total mass of 0.08$-$0.14 M$_{\odot}$.  Assuming that the separation is equal
to the semi-major axis of the system, 2.50$\pm$0.07 AU, the period falls in the range of 10.5 to 13.9 years (Table 3)
and the $\Delta$(RV)$_{max}$ is 5.3$-$7.1 km~s$^{-1}$.  We observed the system in 2001, 2005, and 2006, covering close to half of the
estimated orbital period.  However, the cross-correlation between the 2001 and 2005 spectra yielded a shift of only $-$0.4 km~s$^{-1}$
and between 2001 and 2006 of 1.77 km~s$^{-1}$, indistinguishable within the uncertainty of our RV measurements, especially because the
2006 spectrum was particularly noisy.  Thus we find no evidence for significant orbital motion, implying a longer period or an
unfavorable viewing geometry for the detection of an RV shift, or both.

\subsection{2MASS J05591914-1404488}

The T4.5 dwarf 2M0559 presents an enigmatic case of an over-luminous, extremely low-mass object.  Observers and theorists
alike have speculated (Burgasser 2001; Dahn et al. 2002; Burrows et al. 2006; Dupuy \& Liu 2012) that this
system is an equal mass binary.  Alternatively, there may be fundamental processes at play
in the mid-T dwarf atmospheres that are not yet well-understood.  Specifically, this source is the lynchpin in the J-band
brightening/cloud disruption scenario (Burgasser et al. 2002a).  Zapatero Osorio et al. (2007) estimate limits on possible planetary
mass companions in this system, but such a secondary component would not explain the unusually high brightness.

We obtained four observations of 2M0559 over a time baseline of 6.5 years.  For an age of 1 Gyr, our Monte Carlo simulation (\S 4.1)
indicates a 50\% detection probability for a threshold semi-major axis of 0.13 AU and a 90\% detection probability for a threshold
semi-major axis of 0.003 AU.  The threshold semi-major axis is the separation below which a spectroscopic companion would
be detected with a particular probability.  Burgasser et al. (2003) rule out the presence of a relatively bright companion object closer than 
0$\farcs$09.  At the $\sim$10 pc distance to 2M0559 (Dahn et al. 2002), 0$\farcs$09 corresponds to $\sim$0.9 AU.  Thus, ample
parameter space for a bright binary companion to this object remains unexplored and our confidence in a null result for a 
companion object is only high ($\ge$90\%) for extremely short periods of days or less.  Monitoring this system with extremely
precise RV measurements (see next section) with regular cadence over a considerable time baseline will help to fill in the
potential binary parameter space gap and might also provide insight into the atmospheric properties.

\subsection{The Importance of High Precision RV Measurements}

For spectroscopic binary systems, Figure 7
illustrates the relationships between the primary object's mass, the primary orbital velocity, and the orbital period on the basis of Kepler's third law.
We show results for three distinct values of the mass ratio (q); a circular, edge-on orbit is assumed
for simplicity.  For a system with a primary of mass 0.08, the
substellar limit, and a mass ratio of 1.0, the primary object's RV is $\sim$3.5 km~s$^{-1}$ for a period of 12 years, approximately the shortest period system
among the visual binaries in our sample (Table 3).  With a precision of 2 km~s$^{-1}$, motion of the primary (or the secondary, given a mass ratio of unity)
in such a system is only detectable for very specific phases and viewing angles.  The probability of detection with 2 km~s$^{-1}$
precision increases for shorter-period binaries; however, again, this is only true under certain specialized conditions (\S 4.1).
None of the multi-epoch spectra in our sample of 25 brown dwarf systems reveals more than $\sim$3 km~s$^{-1}$ of RV variability.  Even
for the seven known brown dwarf binaries observed, some with a cadence that regularly sampled a significant fraction of the estimated orbital period,
we were unable to unambiguously detect any RV variability.

Specialized techniques for the
highly precise measurement of small RV shifts, such as those applied to high-resolution K-band spectra
by Blake et al. (2007, 2010), Prato et al. (2008), Konopacky et al. (2010), Bailey et al. (2012), Burgasser et al. (2012), and others,
are required to reliably detect motion in brown dwarf binaries, even for those with orbital periods as short as days.  In their 6-year study of
late-type M and L dwarfs with NIRSPEC on the Keck II telescope, Blake et al. (2010) obtained a precision of 200 m~s$^{-1}$ on slowly rotating
L dwarfs, providing sensitivity to orbital motion of brown dwarf binaries with periods of decades and mass ratios as low as $\sim$10\% (Figure 7),
the upper limit for the detection of giant planetary companions.

In the study described here, even for our sample of 25 systems with zero detections of spectroscopic binaries, it is still not
possible to use the results  to definitively characterize short-period low mass binaries as rare.  The sampling and geometry of such systems
are simply not well-suited to identification with our 2 km~s$^{-1}$ precision and random observing cadence.  Thus as far as it is possible to say
with the extant data, very low mass spectroscopic binaries are not necessarily intrinsically rare, but even with one of the largest samples
available, statistics show (\S 4.1) that 2 km~s$^{-1}$ uncertainties provide relatively weak constraints.

\section{Summary}

We obtained multiple-epoch spectra of a sample of 25 very low-mass field dwarfs, three M dwarfs, sixteen L dwarfs, and six T dwarfs,
between 2000 April and 2011 June to search for
RV variability and spectral evidence for multiple components.  With a precision of $\sim$2 km~s$^{-1}$, we were sensitive to RV
variability at a statistically significant level only in systems with periods of about a day or less, assuming a favorable distribution of
orbital properties and viewing geometries relative to our line of sight.
In none of the systems studied, including the seven known, wide binaries observed, did we detect any RV variability 
$>$3 km~s$^{-1}$.  For over a dozen objects in our sample we present the
first published high-resolution spectra and provide RVs and rotational velocities for the entire sample, either based on
this work or taken from the more precise measurements in Blake et al. (2010).  We show multi-epoch spectral sequences for two
objects of particular interest, Kelu-1 and 2M0937, an L2 and a peculiar T6, respectively.  No significant variations are seen in these
or the other target spectra, some of which boast an exquisite S/N ratio in excess of 100.  

RV measurements of brown dwarfs are important both for the ultimate measurement of brown dwarf
masses (Konopacky et al. 2010) and for the spectroscopic detection of very low-mass, even planetary, companions to presumed single brown dwarfs
(Blake et al. 2010).  The close binary fraction of very low mass systems is highly uncertain (e.g., Bardalez Gagliuffi et al. 2014).
We conclude with the observation that to satisfy these scientific goals requires
high S/N ratio, strategic sampling cadence, and relatively high precision measurements: with the 200 m~s$^{-1}$ precision
of Blake et al., it is possible to detect several-Jupiter mass companions even in orbits of decades (bottom panel, Figure 7).  Long-term
monitoring programs of binary brown dwarfs, and in particular candidate spectroscopic binary brown dwarfs (Bardalez Gagliuffi et al.), with high spectral resolution,
component-resolved spectroscopy (Konopacky et al.), with high spectral resolution unresolved spectroscopy (Burgasser et al. 2012), and with
high-angular resolution imaging (e.g., Radigan et al. 2013), over time scales of days to years are required.  Results of these efforts
will yield component mass measurements with sufficient precision to stringently test models of
brown dwarf structure and evolution, and, in the case of younger systems, formation (e.g., Schaefer et al. 2014).  It is crucial
that RV monitoring programs take advantage of high-precision techniques for a future high-yield science return.

\bigskip
\bigskip

We thank the Keck Observatory OAs and SAs and B. Schaefer, probably all of whom helped with these runs and observations during the
11 year period over which the data were gathered, for their exceptional support.
We are grateful to Q. Konopacky and M. McGovern for assistance with some of the later observing runs.
L.P. thanks O. Franz and L. Wasserman for helpful discussions on orbital dynamics.  We are grateful to the
anonymous referee for comments which improved this manuscript.
Partial support to L.P. for this work was provided by NSF grant AST 04-44017.
This research has benefited from the M, L, T, and Y dwarf compendium housed at DwarfArchives.org.
This work made use of the SIMBAD reference database, the NASA
Astrophysics Data System, and the data products from the Two Micron All
Sky Survey, which is a joint project of the University of Massachusetts
and the Infrared Processing and Analysis Center/California Institute
of Technology, funded by the National Aeronautics and Space
Administration and the National Science Foundation.
Data presented herein were obtained at the W. M. Keck
Observatory, which is operated as a scientific partnership among the California Institute of Technology,
the University of California, and the National Aeronautics and Space Administration. The Observatory
was made possible by the generous financial support of the W. M. Keck Foundation.
The authors recognize and acknowledge the
significant cultural role that the summit of Mauna Kea
plays within the indigenous Hawaiian community.  We are
grateful for the opportunity to conduct observations from this special mountain.

\clearpage

\begin{figure}
\epsscale{0.85}
\figurenum{1}
\centering
\includegraphics[width=5.5in]{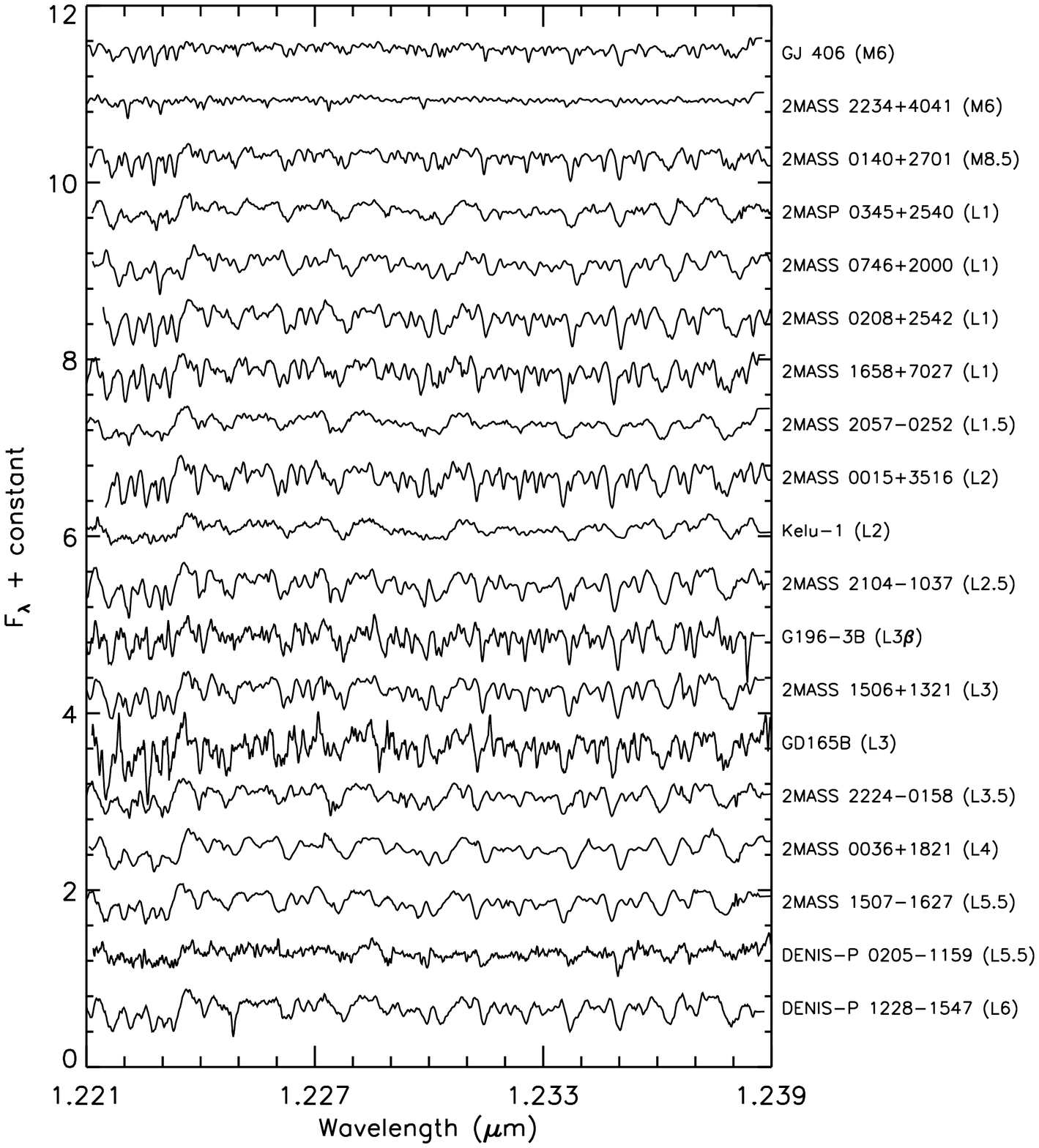}
\caption{The NIRSPEC echelle order 62 M and L dwarf spectral sequence for BDSS RV targets. 
Spectra are corrected for barycentric velocity, normalized, boxcar smoothed to a 3-pixel resolution element, and offset by a constant.
\label{lspecs}}
\end{figure}

\clearpage

\begin{figure}
\epsscale{0.8}
\figurenum{2}
\centering
\includegraphics[width=6in]{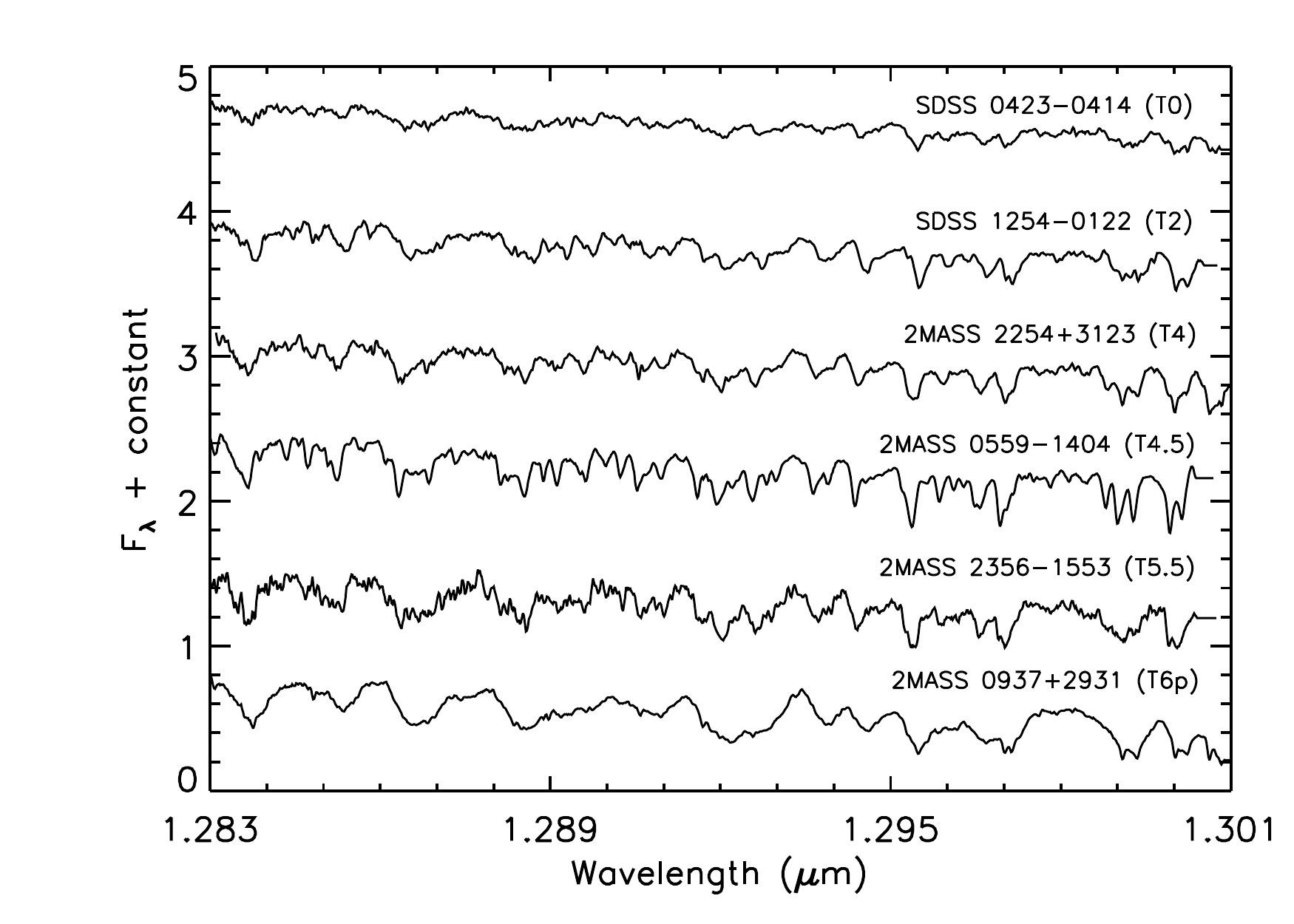}
\caption{The NIRSPEC echelle order 59 T dwarf spectral sequence for BDSS RV targets. 
Spectra are corrected for barycentric velocity, normalized, boxcar smoothed to a 3-pixel resolution element, and offset by a constant.
\label{tspecs}}
\end{figure}

\clearpage

\begin{figure}
\epsscale{0.8}
\figurenum{3}
\centering
\includegraphics[width=6in]{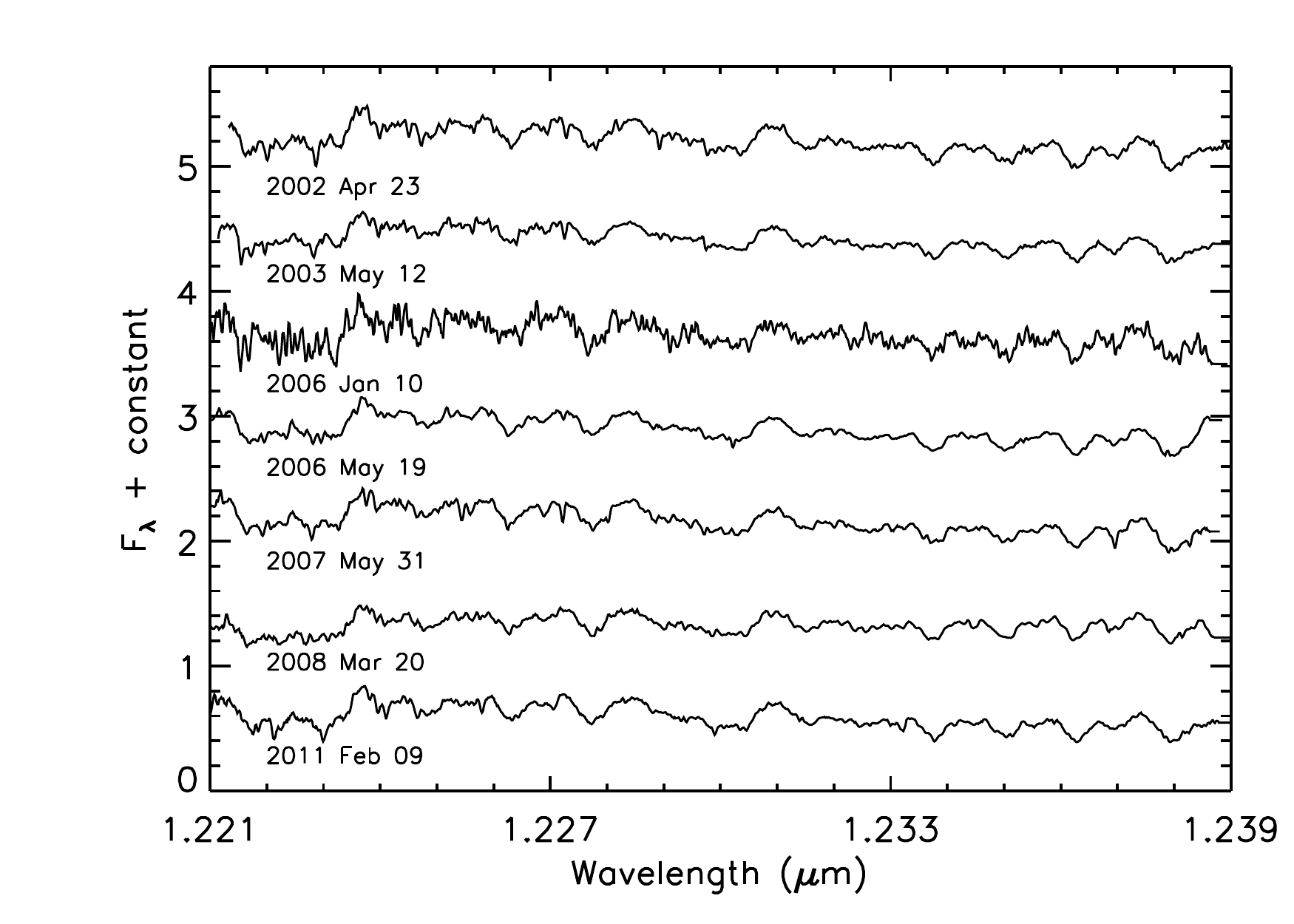}
\caption{Multiple-epoch spectra for NIRSPEC echelle order 62 of the visual binary Kelu-1. 
Spectra are corrected and shifted as in Figures 1 and 2.
We did not measure any significant RV shift over the nearly nine year baseline of the observations (\S 4.1.3).
\label{kelu1}}
\end{figure}

\clearpage

\begin{figure}
\epsscale{0.8}
\figurenum{4}
\centering
\includegraphics[width=6in]{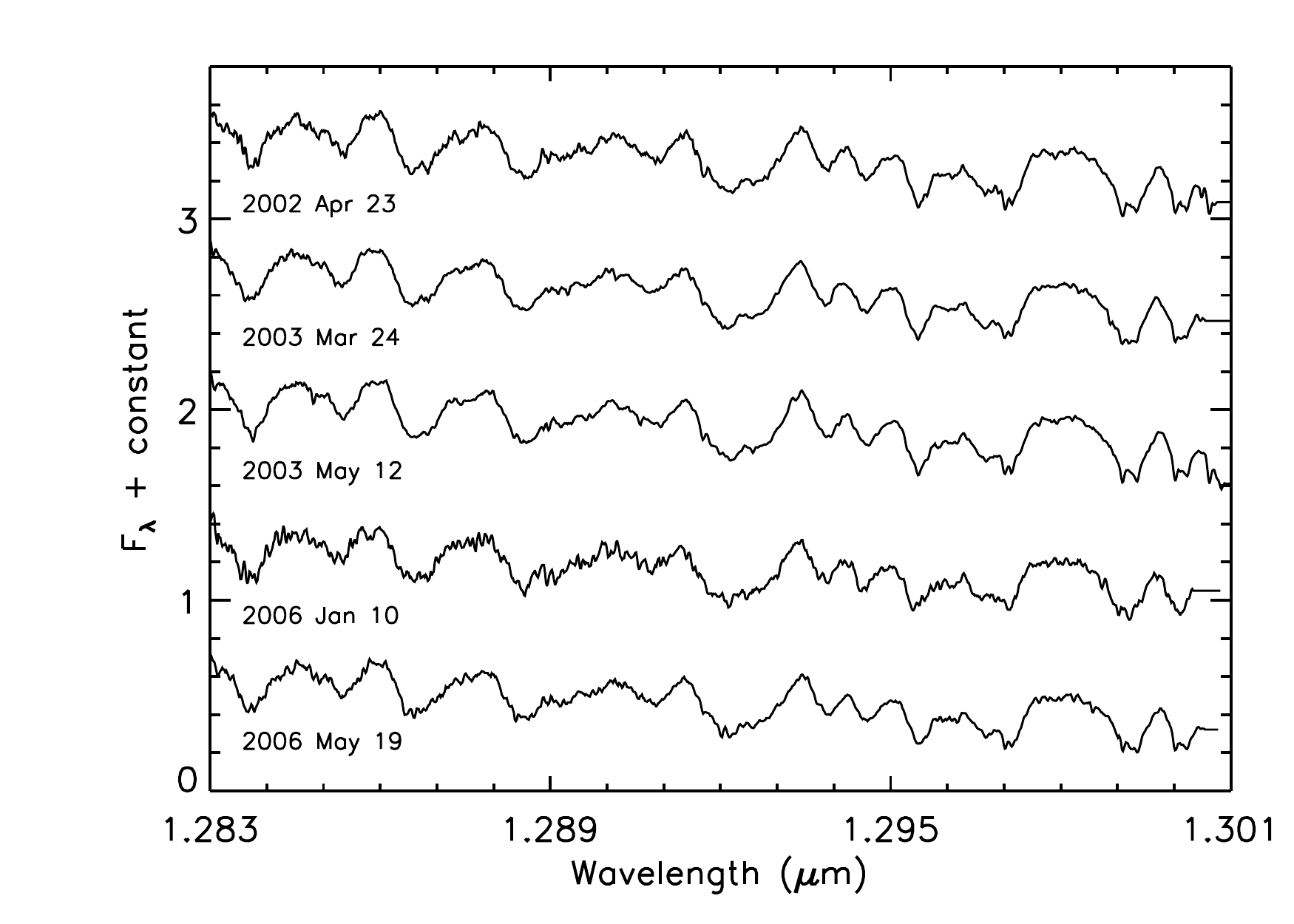}
\caption{Multiple-epoch spectra for NIRSPEC echelle order 59 of the peculiar T6 dwarf 2MASS 0937+2931.
Spectra are corrected and shifted as in Figures 1 and 2.  No significant RV shift or line variability was observed in these spectra.
\label{0937}}
\end{figure}

\clearpage

\begin{figure}
\epsscale{0.9}
\figurenum{5}
\centering
\includegraphics[width=6in]{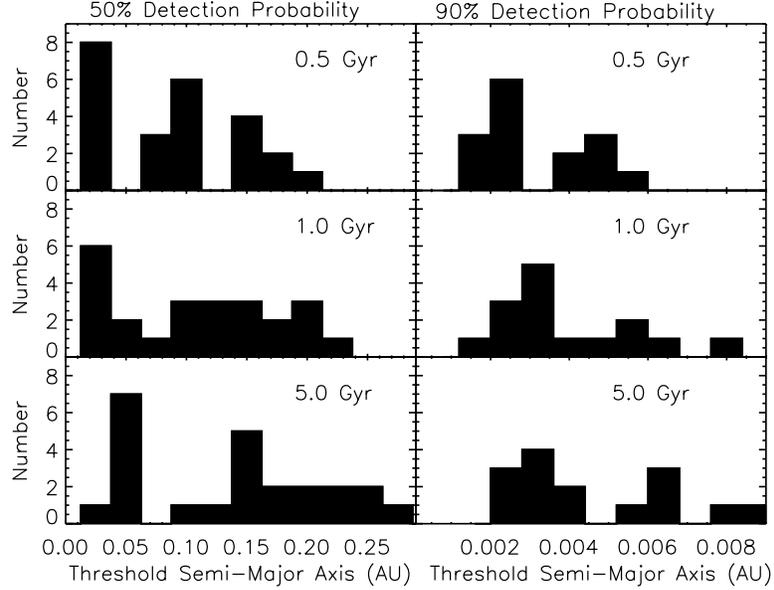}
\caption{Results from our Monte Carlo simulation showing the number of our targets with 50\% (left panels) and 90\% (right panels)
spectroscopic binary detection probability as a function of the threshold semi-major axis (\S 4.3),
given our target properties and RV precision.  Because mass is a function of age, we present results for 3 age bins;
spectral types were converted to mass estimates using models and empirical relations (see text).  The
outliers Kelu-1 and 2M0036, more favorable to detection because
of the combination of their relatively high mass and our high sampling frequency, are not included in the right hand panels but 
ranged between 0.011$-$0.015 AU for a 90\% detection probability.  Kelu-1 is not included in the left panels, either, but its
detection probabilities for the three ages ranged from 0.34$-$0.47 AU for a 50\% detection probability.  In addition to
Kelu-1 and 2M0036, only 15/25 objects had 90\% detection probabilities, thus fewer objects are plotted in the right panels.
Bin sizes are 0.025 AU on the left and 0.0008 AU on the right.
\label{0937}}
\end{figure}

\clearpage

\begin{figure}
\epsscale{0.6}
\figurenum{6}
\centering
\includegraphics[width=5.5in]{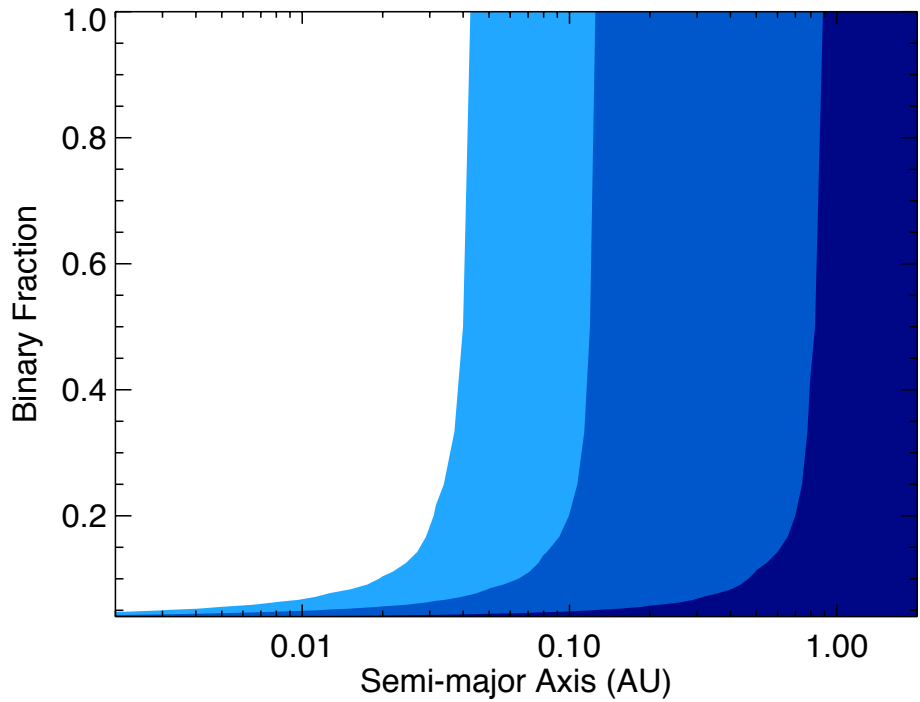}
\caption{Binary fraction detection thresholds as a function of the typical semi-major axis based on the velocity dispersion of our sample. Colored contours correspond to velocity dispersions 1.5, 3, and 5 times higher (from right to left) than the observed dispersion $\sigma$, computed by sampling randomly oriented orbits with the given semi-major axis at the observational epochs of our sample. The most conservative threshold (5$\sigma$) corresponds to semi-major axes of 0.03--0.04~AU.
\label{0937}}
\end{figure}

\clearpage

\begin{figure}
\epsscale{0.6}
\figurenum{7}
\centering
\includegraphics[width=5in]{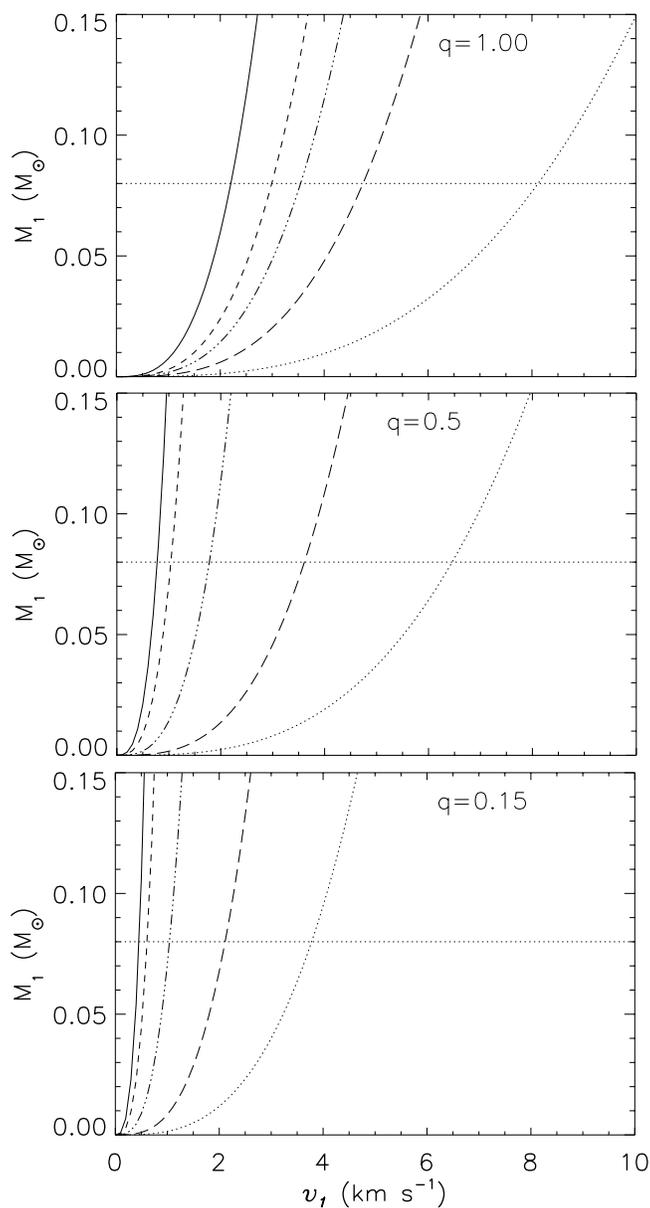}
\caption{Each panel illustrates the relationship in a spectroscopic binary between primary object mass (M$_1$),
primary orbital velocity ($v_1$), and orbital period, curved lines, for 3 sample mass
ratios, q$=$1, 0.5, and 0.15, top to bottom.  Plots for periods of 50 years (solid line), 20 years (short dash line), 12 years (dash-dot line),
5 years (long dash line), and 1 year (dotted line) are shown. The horizontal dotted line demarcates the nominal sub-stellar cut off
for the more massive star in the system.  The orbits are assumed to be circular and edge-on, the most favorable possible orientation for detection.
\label{0937}}
\end{figure}

\clearpage

\pagestyle{empty}

\begin{deluxetable}{lcccccl}
\tablewidth{0pt} \tablecaption{\bf Target List and Observing Log \label{tbl-1}} 
\rotate
\tabletypesize{\tiny}
\tablehead{ \colhead{$ $} &
\colhead{$\alpha$} & \colhead{$\delta$} &
\colhead{Spectral} & \colhead{$J$} & Discovery & \colhead{UT Dates}\\
\colhead{Object} & \colhead{(J2000.0)} & \colhead{(J2000.0)} &
\colhead{Type\tablenotemark{a}} & \colhead{(mag)} & Reference & \colhead{of Observation}\\
\colhead{(1)} & \colhead{(2)} & \colhead{(3)} &
\colhead{(4)} & \colhead{(5)} & \colhead{(6)} & \colhead{(7)}} 
\startdata
GJ 406 (Wolf 359)  & 10 56 28.9 & 07 00 53 & M$_o$6 & 7.09$\pm$0.02 & 1 & 2002 Apr 23, 2006 Jan 11, May 20 \\
2MASS J22344161+4041387\tablenotemark{b} & 22 34 41.6 & 40 41 39 & M$_o$6 & 12.57$\pm$0.02 & 2 & 2006 Oct 6, 2007 May 29, 2009 Nov 8 \\
2MASS J01400263+2701505 & 01 40 02.6 & 27 01 51 & M$_o$8.5 & 12.49$\pm$0.02 & 3 & 2000 Dec 4, Dec 6, 2002 June 23 \\
2MASP J0345432+254023 & 03 45 43.2 & 25 40 23 & L1 & 14.00$\pm$0.03 & 4 & 2000 Dec 4, Dec 6, 2006 Jan 11 \\
2MASS J07464256+2000321\tablenotemark{b}  & 07 46 42.6 & 20 00 32 & L1 & 11.76$\pm$0.02 & 5 & 2002 Jan 1, 2006 Jan 10 \\
2MASS J02081833+2542533 & 02 08 18.3 & 25 42 53 & L$_o$1 & 13.99$\pm$0.03 & 6 & 2000 Jul 25, Jul 29, 2008 Dec 6 \\
2MASS J16580380+7027015 & 16 58 03.8 & 70 27 02 & L$_o$1 & 13.29$\pm$ 0.02 & 3 & 2000 Jul 25, 2007 May 29 \\
2MASS J20575409-0252302 & 20 57 54.1 & -02 52 30 & L1.5 & 13.12$\pm$0.02 & 2 & 2000 Jul 25, 2003 Jul 20, 2007 May 29 \\
2MASS J00154476+3516026	& 00 15 44.8  & 35 16 03  & L$_o$2 & 13.87$\pm$0.03 & 6 & 2000 Jul 25, 2008 Dec 5 \\
Kelu-1\tablenotemark{b} & 13 05 40.2 &-25 41 06 & L$_o$2 & 13.41$\pm$0.03 & 7 & 2002 Apr 23, 2003 May 12, 2006 Jan 10, May 19, 2007 May 31, 2008 Mar 20, 2011 Feb 9 \\
2MASS J21041491-1037369 & 21 04 14.9 & -10 37 37 & L$_o$2.5 & 13.84$\pm$0.03 & 2 & 2007 May 30, 2009 Nov 8 \\
G196-3B      & 10 04 20.7 & 50 23 00 & L$_o$3$\beta$\tablenotemark{d} & 14.83$\pm$0.05 & 8 & 2002 Apr 23, 2006 Jan 11, May 20, 2008 Mar 21 \\
2MASS J15065441+1321060 & 15 06 54.4 & 13 21 06 & L$_o$3 & 13.37$\pm$0.02 & 3 & 2000 Jul 25, Jul 28, 2007 May 30, 2008 Mar 21 \\
GD165B             & 14 24 39.1 & 09 17 10 & L3  & 15.69$\pm$0.08 & 9 & 2002 Apr 23, 2003 May 13, 2006 May 20, 2011 Feb 5, 9 \\
2MASS J22244381-0158521 & 22 24 43.8 & -01 58 52 & L3.5 & 14.07$\pm$0.03 & 6 & 2000 Jul 25, 2007 May 30 \\
2MASS J00361617+1821104 & 00 36 16.2 & 18 21 10 & L4 & 12.47$\pm$0.03 & 5 & 2000 Jul 25, Jul 29, Dec 4, Dec 5, Dec 6, 2001 Dec 30, 2005 Dec 11 \\
2MASS J15074769-1627386\tablenotemark{c} & 15 07 47.7 &-16 27 39 & L5.5 & 12.83$\pm$0.03 & 5 & 2000 Apr 25, Jul 25, Jul 28, 2007 May 29, 2008 Mar 21 \\
DENIS-P J0205.4-1159\tablenotemark{b} & 02 05 29.4 &-11 59 30 & L5.5 & 14.59$\pm$0.03 & 10 & 2001 Oct 9, Dec 29, 2006 Jan 10 \\
DENIS-P J1228.2-1547\tablenotemark{b} & 12 28 15.2 & -15 47 35 & L6 & 14.38$\pm$0.03 & 10 & 2007 May 30, 2011 Feb 5, June 8 \\
SDSSp J042348.57-041403.5\tablenotemark{b} & 04 23 48.6 &-04 14 04 & T0  & 14.47$\pm$0.03  & 11 & 2001 Oct 9, 2005 Dec 11, 2006 Jan 10 \\
SDSSp J125453.90-012247.4 & 12 54 53.9 &-01 22 47 & T2  & 14.89$\pm$0.04 & 12 & 2003 May 14, 2007 May 31 \\
2MASS J22541892+3123498 & 22 54 18.9 & 31 23 50 & T4 & 15.26$\pm$0.05 & 13 & 2003 Aug 10, 2005 Jul 19 \\
2MASS J05591914-1404488 & 05 59 19.1 &-14 04 49 & T4.5 & 13.80$\pm$0.02  & 14 & 2001 Oct 9, Dec 29, 2006 Jan 11, 2008 Mar 19 \\
2MASS J23565477-1553111 & 23 56 54.8 & -15 53 11 & T5.5 & 15.82$\pm$0.06  & 13 & 2005 Jul 19, Dec 10 \\
2MASS J09373487+2931409 & 09 37 34.9 & 29 31 41 & T6p & 14.65$\pm$0.04 & 13 & 2002 Apr 23, 2003 Mar 24, May 12, 2006 Jan 10, May 19
\enddata
\tablecomments{References: 1) Wolf (1919); 2) Cruz et al. (2003); 3) Gizis et al. (2000); 4) Kirkpatrick et al. (1997); 5) Reid et al. (2000); 6) Kirkpatrick(2000); 7) Ruiz et al. (1997);\\
8) Reboloet al. (1998); 9) Becklin \& Zuckerman (1988); 10) Delfosse et al. (1997); 11) Geballe et al. (2002); 12) Leggett et al. (2000); 13) Burgasser et al. (2002b); 14) Burgasser et al. (2000).}
\tablenotetext{a}{Spectral types are from DwarfArchives.org; IR type is provided unless only optical type is available, indicated with $_o$ subscript.}
\tablenotetext{b}{Binary system.}
\tablenotetext{c}{RV variable according to Blake et al. (2010).}
\tablenotetext{d}{The $\beta$ designation indicates intermediate surface gravity (Cruz, Kirkpatrick, Burgasser 2009).}

\end{deluxetable}

\clearpage

\pagestyle{empty}

\begin{deluxetable}{lccccl}
\tablewidth{0pt} \tablecaption{\bf Radial and Rotational Velocities \label{tbl-2}} 
%\rotate
\tabletypesize{\small}
\tablehead{\colhead{$ $} &
\colhead{N}  & \colhead{$\Delta$T} & \colhead{$<RV>$}
& \colhead{$v \sin i$} & \colhead{References}\\
\colhead{Object} & \colhead{Obs.} & \colhead{(days)} & \colhead{(km s$^{-1}$)}
& \colhead{(km s$^{-1}$)} & \colhead{$<RV>$, $v \sin i$}\\
\colhead{(1)} & \colhead{(2)} & \colhead{(3)} &
\colhead{(4)} & \colhead{(5)} & \colhead{(6)}} 
\startdata
GJ 406 & 3 & 1488 & 19$\pm$2 & $\le$8$\pm$2 & Rice et al. (2010)\\
2M2234$+$40\tablenotemark{a} & 3  & 2166 & $-$10$\pm$2 & 17$\pm$2 & This work, Rice et al. (2010)\\
2M0140$+$27 & 3 & 566 & 9$\pm$2 & 11$\pm$2 & Rice et al. (2010)\\
2M0345$+$25 & 3 & 1864 & 6$\pm$3 &  30$\pm$5 &  This work \\
2M0746$+$20\tablenotemark{a} & 2 & 1470 &  52.37$\pm$0.06 & 32.72$\pm$0.56 & Blake et al. (2010)\\ 
2M0208$+$25 & 3 & 3056 & 20$\pm$2 &  12$\pm$2 &  This work \\
2M1658$+$70 & 2 & 499 & $-$25.60$\pm$0.12 & 12.26$\pm$0.76 & Blake et al. (2010)\\
2M2057$-$02& 3 & 499 & $-$24.68$\pm$0.43 & 60.56$\pm$0.37 & Blake et al. (2010)\\
2M0015$+$35	& 2 & 3054 & $-$37.35$\pm$0.16 & 10.23$\pm$2.55 & Blake et al. (2010)\\
Kelu-1\tablenotemark{a}  & 7 & 3214 & 6.37$\pm$0.35 & 68.88$\pm$2.60 & Blake et al. (2010)\\
2M2104$-$10 & 2 & 893 & $-$21.09$\pm$0.12 & 23.44$\pm$0.23 & Blake et al. (2010)\\
G196-3B            & 4 & 2159 & $-$2$\pm$2 &  10$\pm$2 & This work, Mohanty \& Basri (2003)\\
2M1506$+$13 & 4 &  796 & $-$0.68$\pm$0.11 & 11.39$\pm$0.94 & Blake et al. (2010)\\
GD165B             & 5 & 3214 &  $-$29$\pm$2 & 18$\pm$2  & This work, Mohanty \& Basri (2003)\\
2M2224$-$01 & 2 & 2500 & $-$37.55$\pm$0.09 & 25.49$\pm$0.41 & Blake et al. (2010)\\
2M0036$+$18 & 7 & 1965 & 19.02$\pm$0.15 & 35.12$\pm$0.57 & Blake et al. (2010)\\
2M1507$-$16 & 5 & 2887 & $-$39.85$\pm$0.05 & 21.27$\pm$1.86 & Blake et al. (2010)\\
DENIS 0205\tablenotemark{a} & 3 & 1554 & 7$\pm$2 & 22$\pm$5 & This work, Basri et al. (2000)\\
DENIS 1228\tablenotemark{a} & 3 & 1347 & 2$\pm$2 & 22$\pm$3 & This work, Basri et al. (2000)\\
SDSS 0423$-$04\tablenotemark{a} & 3 & 1554 & 28$\pm$2 & 60$\pm$10 &  This work \\
SDSS 1254$-$01 & 2 & 1478 & 4$\pm$3 &  15$\pm$5 &  This work \\
2M2254$+$31 & 2 & 709 & 14$\pm$3 &  15$\pm$5 &  This work \\
2M0559$-$14 & 4 & 2353 & $-$9$\pm$3 & 10$\pm$5 &  This work \\
2M2356$-$15 & 2 & 144 & 19$\pm$3 & 15$\pm$5 &  This work \\
2M0937$+$29 & 5 & 1487 & $-$5$\pm$3 & 60$\pm$10 &   This work \\
\enddata
\tablenotetext{a}{Binary system.}\\
%\tablenotetext{b}{yyyyyyy}

\end{deluxetable}

\clearpage

\pagestyle{empty}

\begin{deluxetable}{llllcccl}
\tablewidth{0pt} \tablecaption{\bf Observed and Estimated Binary Properties \label{tbl-3}} 
\rotate
\tabletypesize{\small}
\tablehead{\colhead{$ $} &
\colhead{$M_{total}$}  & \colhead{Distance} & \colhead{Separation}
& \colhead{Semi-Major Axis\tablenotemark{a}} & \colhead{Period\tablenotemark{a}} & \colhead{$\Delta$(RV)$_{max}$\tablenotemark{a}} & \colhead{$  $}\\
\colhead{Object} & \colhead{($M_{\odot}$)} & \colhead{(pc)} & \colhead{($''$)}
& \colhead{(AU)} & \colhead{(years)} & \colhead{(km~s$^{-1}$)} & \colhead{Reference}\\
\colhead{(1)} & \colhead{(2)} & \colhead{(3)} &
\colhead{(4)} & \colhead{(5)} & \colhead{(6)} & \colhead{(7)} & \colhead{(8)}} 
\startdata
2M2234$+$40 & $0.20^{+0.11}_{-0.06}$ & $325^{+72}_{-50}$ & $0.1582\pm0.0003$ & $51^{+12}_{-8}$ & $824^{+510}_{-310}$ & $1.9^{+0.6}_{-0.5}$ & Allers et al. (2009)\\
2M0746$+$20\tablenotemark{b} & $0.151\pm0.003$ & $12.21\pm0.05$\tablenotemark{d} & $0.2373^{+0.0015}_{-0.0040}$\tablenotemark{c} & $2.90^{+0.03}_{-0.06}$ & $12.71\pm0.07$ & $6.8\pm0.1$\tablenotemark{e} & Konopacky et al. (2010)\\
Kelu-1 & $0.115\pm0.014$ & $18.7\pm0.7$\tablenotemark{d} & $0.298\pm0.003$ & $5.6\pm0.3$ & $39\pm5$ & $4.3\pm0.4$ & Gelino et al. (2006) \\
DENIS 0205 & 0.15 & $19.76\pm0.57$\tablenotemark{d} & 0.35\tablenotemark{f} & 6.9 & 47 & 4.4 & Bouy et al. (2006)\\
DENIS 1228 & 0.135 & $20.24\pm0.08$\tablenotemark{d} & 0.32\tablenotemark{c} & 6.41 & 44.2 & 4.3 & Brandner et al. (2004)\\
SDSS 0423$-$04 & $0.08-0.14$ & $15.2\pm0.4$\tablenotemark{g} & $0.1642\pm0.0017$ & $2.50\pm0.07$ & $10.5-13.9$ & $5.3-7.1$ & Burgasser et al. (2005)\\
\enddata
\tablenotetext{a}{Assuming eccentricity is 0 and inclination is 90 degrees.}
\tablenotetext{b}{All parameters except for $\Delta$(RV)$_max$ from Konopacky et al. 2010.}
\tablenotetext{c}{Semi-major axis ($''$).}
\tablenotetext{d}{Dahn et al. (2002).}
\tablenotetext{e}{Average $v_{orbital}$, modulo inclination of 138 degrees; consistent with Konopacky et al. (2010) fit to $K_1+K_2$.}
\tablenotetext{f}{Average observed separation from Bouy et al. (2006).}
\tablenotetext{g}{Vrba et al. (2004).}

\end{deluxetable}


\begin{thebibliography}{}

\bibitem[Abt \& Levy(1976)]{1976ApJS...30..273A} Abt, H.~A., \& Levy, S.~G.\ 1976, \apjs, 30, 273

\bibitem[Allers et al.(2009)]{2009ApJ...697..824A} Allers, K.~N., Liu, 
M.~C., Shkolnik, E., et al.\ 2009, \apj, 697, 824

\bibitem[Armitage 
\& Bonnell(2002)]{2002MNRAS.330L..11A} Armitage, P.~J., \& Bonnell, I.~A.\ 2002, \mnras, 330, L11

\bibitem[Bailey et al.(2012)]{2012ApJ...749...16B} Bailey, J.~I., III, 
White, R.~J., Blake, C.~H., et al.\ 2012, \apj, 749, 16

\bibitem[Bakos et al.(2009)]{2009ApJ...707..446B} Bakos, G.~{\'A}., Howard, 
A.~W., Noyes, R.~W., et al.\ 2009, \apj, 707, 446

\bibitem[Bardalez Gagliuffi et al.(2014)]{2014ApJ...794..143B} Bardalez 
Gagliuffi, D.~C., Burgasser, A.~J., Gelino, C.~R., et al.\ 2014, \apj, 794, 143

\bibitem[Basri \& Mart{\'{\i}}n(1999)]{1999AJ....118.2460B} Basri, G., \& Mart{\'{\i}}n, E.~L.\ 1999, \aj, 118, 2460 

\bibitem[Basri et al.(2000)]{2000ApJ...538..363B} Basri, G., Mohanty, S., 
Allard, F., et al.\ 2000, \apj, 538, 363

\bibitem[Basri \& Reiners(2006)]{2006AJ....132..663B} Basri, G., \& Reiners, A.\ 2006, \aj, 132, 663

\bibitem[Becklin \& Zuckerman(1988)]{1988Natur.336..656B} Becklin, E.~E., \& Zuckerman, B.\ 1988, \nat, 336, 656

\bibitem[Bennett et al.(2008)]{2008ApJ...684..663B} Bennett, D.~P., Bond, 
I.~A., Udalski, A., et al.\ 2008, \apj, 684, 663

\bibitem[Blake et al.(2007)]{bla07} Blake, C.~H., Charbonneau, D., White, R.~J.,
Marley, M.~S., \& Saumon, D.\ 2007, \apj, 666, 1198

\bibitem[Blake et al.(2008)]{bla08} Blake, C.~H., Charbonneau, D., White, R.~J.,
Torres, G., Marley, M.~S., \& Saumon, D.\ 2008, \apjl, 678, L125 

\bibitem[Blake et al.(2010)]{2010ApJ...723..684B} Blake, C.~H., 
Charbonneau, D., \& White, R.~J.\ 2010, \apj, 723, 684

\bibitem[Bouy et al.(2003)]{2003AJ....126.1526B} Bouy, H., Brandner, W., 
Mart{\'{\i}}n, E.~L., et al.\ 2003, \aj, 126, 1526

\bibitem[Bouy et al.(2005)]{2005AJ....129..511B} Bouy, H., Mart{\'{\i}}n, 
E.~L., Brandner, W., \& Bouvier, J.\ 2005, \aj, 129, 511

\bibitem[Bouy et al.(2006)]{2006A&A...451..177B} Bouy, H., Mart{\'{\i}}n, E.~L., Brandner, W., et al.\ 2006, \aap, 451, 177

\bibitem[Brandner et al.(2004)]{2004A&A...428..205B} Brandner, W., Mart{\'{\i}}n, E.~L.,
Bouy, H., et al.\ 2004, \aap, 428, 205

\bibitem[Burgasser et al.(2000)]{2000AJ....120.1100B} Burgasser, A.~J., 
Wilson, J.~C., Kirkpatrick, J.~D., et al.\ 2000, \aj, 120, 1100

\bibitem[Burgasser(2001)]{2001PhDT.......116B} Burgasser, A.~J.\ 2001, 
Ph.D.~Thesis

\bibitem[Burgasser et al.(2002a)]{2002aApJ...571L.151B} Burgasser, A.~J., 
Marley, M.~S., Ackerman, A.~S., et al.\ 2002a, \apjl, 571, L151

\bibitem[Burgasser et al.(2002b)]{2002bApJ...564..421B} Burgasser, A.~J., 
Kirkpatrick, J.~D., Brown, M.~E., et al.\ 2002b, \apj, 564, 421

\bibitem[Burgasser et al.(2003)]{2003ApJ...586..512B} Burgasser, A.~J., 
Kirkpatrick, J.~D., Reid, I.~N., et al.\ 2003, \apj, 586, 512

\bibitem[Burgasser et al.(2005)]{2005ApJ...634L.177B} Burgasser, A.~J., 
Reid, I.~N., Leggett, S.~K., et al.\ 2005, \apjl, 634, L177

\bibitem[Burgasser et al.(2008)]{2008ApJ...681..579B} Burgasser, A.~J., 
Liu, M.~C., Ireland, M.~J., Cruz, K.~L., \& Dupuy, T.~J.\ 2008, \apj, 681, 579

\bibitem[Burgasser et al.(2012)]{2012ApJ...757..110B} Burgasser, A.~J., 
Luk, C., Dhital, S., et al.\ 2012, \apj, 757, 110

\bibitem[Burgasser et al.(2014)]{2014arXiv1410.4288B} Burgasser, A.~J., 
Gillon, M., Melis, C., et al.\ 2014, arXiv:1410.4288

\bibitem[Burrows et al.(2001)]{2001RvMP...73..719B} Burrows, A., Hubbard, 
W.~B., Lunine, J.~I., \& Liebert, J.\ 2001, Reviews of Modern Physics, 73, 719

\bibitem[Burrows et al.(2006)]{2006ApJ...640.1063B} Burrows, A., Sudarsky, 
D., \& Hubeny, I.\ 2006, \apj, 640, 1063

\bibitem[Crockett et al.(2012)]{2012ApJ...761..164C} Crockett, C.~J., 
Mahmud, N.~I., Prato, L., et al.\ 2012, \apj, 761, 164

\bibitem[Cruz et al.(2003)]{2003AJ....126.2421C} Cruz, K.~L., Reid, I.~N., 
Liebert, J., Kirkpatrick, J.~D., \& Lowrance, P.~J.\ 2003, \aj, 126, 2421

\bibitem[Cruz et al.(2009)]{2009AJ....137.3345C} Cruz, K.~L., Kirkpatrick, 
J.~D., \& Burgasser, A.~J.\ 2009, \aj, 137, 3345

\bibitem[Dahn et al.(2002)]{2002AJ....124.1170D} Dahn, C.~C., Harris, 
H.~C., Vrba, F.~J., et al.\ 2002, \aj, 124, 1170

\bibitem[Dahn et al.(2008)]{2008ApJ...686..548D} Dahn, C.~C., Harris, 
H.~C., Levine, S.~E., et al.\ 2008, \apj, 686, 548

\bibitem[Delfosse et al.(1997)]{1997A&A...327L..25D} Delfosse, X., Tinney, C.~G., Forveille, T., et al.\ 1997, \aap, 327, L25

\bibitem[Dupuy et al.(2010)]{2010ApJ...721.1725D} Dupuy, T.~J., Liu, M.~C., 
Bowler, B.~P., et al.\ 2010, \apj, 721, 1725

\bibitem[Dupuy \& Liu(2011)]{2011ApJ...733..122D} Dupuy, T.~J., \& Liu, M.~C.\ 2011, \apj, 733, 122

\bibitem[Dupuy \& Liu(2012)]{2012ApJS..201...19D} Dupuy, T.~J., \& Liu, M.~C.\ 2012, \apjs, 201, 19

\bibitem[Duquennoy \& Mayor(1991)]{1991A&A...248..485D} Duquennoy, A., \& Mayor, M.\ 1991, \aap, 248, 485

\bibitem[Geballe et al.(2002)]{2002ApJ...564..466G} Geballe, T.~R., Knapp, 
G.~R., Leggett, S.~K., et al.\ 2002, \apj, 564, 466

\bibitem[Gelino et al.(2006)]{2006PASP..118..611G} Gelino, C.~R., Kulkarni, 
S.~R., \& Stephens, D.~C.\ 2006, \pasp, 118, 611

\bibitem[Gizis et al.(2000)]{2000AJ....120.1085G} Gizis, J.~E., Monet, 
D.~G., Reid, I.~N., et al.\ 2000, \aj, 120, 1085

\bibitem[Golimowski et al.(2004)]{2004AJ....128.1733G} Golimowski, D.~A., 
Henry, T.~J., Krist, J.~E., et al.\ 2004, \aj, 128, 1733

\bibitem[Grether \& Lineweaver(2006)]{2006ApJ...640.1051G} Grether, D., \& Lineweaver, C.~H.\ 2006, \apj, 640, 1051

\bibitem[Joergens(2008)]{2008A&A...492..545J} Joergens, V.\ 2008, \aap, 492, 545

\bibitem[Joergens et al.(2010)]{2010A&A...521A..24J} Joergens, V., M{\"u}ller, A., \& Reffert, S.\ 2010, \aap, 521, A24 

\bibitem[Joergens et 
al.(2013)]{2013A&A...558L...7J} Joergens, V., Bonnefoy, M., Liu, Y., et al.\ 2013, \aap, 558, LL7

\bibitem[Kirkpatrick et al.(1997)]{1997ApJ...476..311K} Kirkpatrick, J.~D., 
Beichman, C.~A., \& Skrutskie, M.~F.\ 1997, \apj, 476, 311

\bibitem[Kirkpatrick et al.(2000)]{2000AJ....120..447K} Kirkpatrick, J.~D., 
Reid, I.~N., Liebert, J., et al.\ 2000, \aj, 120, 447

\bibitem[Koerner et al.(1999)]{1999ApJ...526L..25K} Koerner, D.~W., 
Kirkpatrick, J.~D., McElwain, M.~W., \& Bonaventura, N.~R.\ 1999, \apjl, 526, L25

\bibitem[Konopacky et al.(2010)]{2010ApJ...711.1087K} Konopacky, Q.~M., 
Ghez, A.~M., Barman, T.~S., et al.\ 2010, \apj, 711, 1087

\bibitem[Leggett et al.(2000)]{2000ApJ...536L..35L} Leggett, S.~K., 
Geballe, T.~R., Fan, X., et al.\ 2000, \apjl, 536, L35

\bibitem[Liu \& Leggett(2005)]{2005ApJ...634..616L} Liu, M.~C., \& Leggett, S.~K.\ 2005, \apj, 634, 616

\bibitem[Lodieu et al.(2007)]{2007MNRAS.379.1423L} Lodieu, N., Pinfield, 
D.~J., Leggett, S.~K., et al.\ 2007, \mnras, 379, 1423

\bibitem[Looper et al.(2010)]{2010AJ....140.1486L} Looper, D.~L., 
Bochanski, J.~J., Burgasser, A.~J., et al.\ 2010, \aj, 140, 1486

\bibitem[Mart{\'{\i}}n et al.(1999)]{1999AJ....118.2466M} Mart{\'{\i}}n, 
E.~L., Delfosse, X., Basri, G., et al.\ 1999, \aj, 118, 2466

\bibitem[Mazeh et al.(2003)]{2003ApJ...599.1344M} Mazeh, T., Simon, M., 
Prato, L., Markus, B., \& Zucker, S.\ 2003, \apj, 599, 1344

\bibitem[McGovern et al.(2004)]{2004ApJ...600.1020M} McGovern, M.~R., 
Kirkpatrick, J.~D., McLean, I.~S., Burgasser, A.~J., Prato, L., 
\& Lowrance, P.~J.\ 2004, \apj, 600, 1020 

\bibitem[McLean et al.(1998)]{mclean98} McLean, I. S., et al. 1998,
SPIE, 3354, 566

\bibitem[McLean et al.(2000)]{mclean00} McLean, I. S., Graham, J. R.,
Becklin, E. E., Figer, D. F., Larkin, J. E., Levenson, N. A., \& Teplitz,
H. I. 2000, SPIE, 4008, 1048

\bibitem[McLean et al.(2001)]{2001ApJ...561L.115M} McLean, I.~S., Prato, 
L., Kim, S.~S., Wilcox, M.~K., Kirkpatrick, J.~D., 
\& Burgasser, A.\ 2001, \apjl, 561, L115 

\bibitem[McLean et al.(2003)]{2003ApJ...596..561M} McLean, I.~S., McGovern, 
M.~R., Burgasser, A.~J., Kirkpatrick, J.~D., Prato, L., 
\& Kim, S.~S.\ 2003, \apj, 596, 561 

\bibitem[McLean et al.(2007)]{2007ApJ...658.1217M} McLean, I.~S., Prato, 
L., McGovern, M.~R., Burgasser, A.~J., Kirkpatrick, J.~D., Rice, E.~L., 
\& Kim, S.~S.\ 2007, \apj, 658, 1217

\bibitem[Metchev \& Hillenbrand(2009)]{2009ApJS..181...62M} Metchev, S.~A., \&
Hillenbrand, L.~A.\ 2009, \apjs, 181, 62

\bibitem[Mohanty \& Basri(2003)]{2003ApJ...583..451M} Mohanty, S., \& Basri, G.\ 2003, \apj, 583, 451

\bibitem[Prato et al.(2008)]{2008ApJ...687L.103P} Prato, L., Huerta, M., 
Johns-Krull, C.~M., et al.\ 2008, \apjl, 687, L103

\bibitem[Radigan et al.(2013)]{2013ApJ...778...36R} Radigan, J., 
Jayawardhana, R., Lafreni{\`e}re, D., et al.\ 2013, \apj, 778, 36

\bibitem[Raghavan et al.(2010)]{2010ApJS..190....1R} Raghavan, D., et al.\ 
2010, \apjs, 190, 1 

\bibitem[Rebolo et al.(1998)]{1998Sci...282.1309R} Rebolo, R., Zapatero 
Osorio, M.~R., Madruga, S., et al.\ 1998, Science, 282, 1309

\bibitem[Reid et al.(2000)]{2000AJ....119..369R} Reid, I.~N., Kirkpatrick, 
J.~D., Gizis, J.~E., et al.\ 2000, \aj, 119, 369

\bibitem[Reid et al.(2006)]{2006AJ....132..891R} Reid, I.~N., Lewitus, E., 
Allen, P.~R., Cruz, K.~L., \& Burgasser, A.~J.\ 2006, \aj, 132, 891

\bibitem[Reipurth 
\& Clarke(2001)]{2001AJ....122..432R} Reipurth, B., \& Clarke, C.\ 2001, \aj, 122, 432

\bibitem[Rice et al.(2010)]{2010ApJS..186...63R} Rice, E.~L., Barman, T., 
Mclean, I.~S., Prato, L., \& Kirkpatrick, J.~D.\ 2010, \apjs, 186, 63 

\bibitem[Ruiz et al.(1997)]{1997ApJ...491L.107R} Ruiz, M.~T., Leggett, 
S.~K., \& Allard, F.\ 1997, \apjl, 491, L107

\bibitem[Schaefer et al.(2014)]{2014AJ....147..157S} Schaefer, G.~H., 
Prato, L., Simon, M., \& Patience, J.\ 2014, \aj, 147, 157

\bibitem[Shkolnik et al.(2012)]{2012ApJ...758...56S} Shkolnik, E.~L., 
Anglada-Escud{\'e}, G., Liu, M.~C., et al.\ 2012, \apj, 758, 56

\bibitem[Siegler et al.(2003)]{2003ApJ...598.1265S} Siegler, N., Close, 
L.~M., Mamajek, E.~E., \& Freed, M.\ 2003, \apj, 598, 1265

\bibitem[Simon et al.(2006)]{2006ApJ...644.1183S} Simon, M., Bender, C., 
\& Prato, L.\ 2006, \apj, 644, 1183

\bibitem[Stassun et al.(2006)]{2006Natur.440..311S} Stassun, K.~G., 
Mathieu, R.~D., \& Valenti, J.~A.\ 2006, \nat, 440, 311

\bibitem[Stephens et al.(2009)]{2009ApJ...702..154S} Stephens, D.~C., 
Leggett, S.~K., Cushing, M.~C., et al.\ 2009, \apj, 702, 154

\bibitem[Stumpf et al.(2008)]{2008arXiv0811.0556S} Stumpf, M.~B., Brandner, 
W., Henning, T., et al.\ 2008, arXiv:0811.0556

\bibitem[Sumi et al.(2011)]{2011Natur.473..349S} Sumi, T., Kamiya, K., 
Bennett, D.~P., et al.\ 2011, \nat, 473, 349

\bibitem[Vrba et al.(2004)]{2004AJ....127.2948V} Vrba, F.~J., Henden, 
A.~A., Luginbuhl, C.~B., et al.\ 2004, \aj, 127, 2948

\bibitem[Wolf(1919)]{1919VeHei...7..195W} Wolf, M.\ 1919, 
Veroeffentlichungen der Badischen Sternwarte zu Heidelberg, 7, 195

\bibitem[Zapatero Osorio et al.(2004)]{2004ApJ...615..958Z} Zapatero 
Osorio, M.~R., Lane, B.~F., Pavlenko, Y., et al.\ 2004, \apj, 615, 958

\bibitem[Zapatero Osorio et al.(2007)]{2007ApJ...666.1205Z} Zapatero 
Osorio, M.~R., Mart{\'{\i}}n, E.~L., B{\'e}jar, V.~J.~S., et al.\ 2007, \apj, 666, 1205

\end{thebibliography}
\end{document}